# Siren: Advancing Cybersecurity Through Deception and Adaptive Analysis

Ananthanarayanan Samhruth[1], Kulathumani Girish[2], and Narayanan Ganesh[3]

School of Computer Science and Engineering, Vellore Institute of Technology, Chennai
samhruth.ananth2021@vitstudent.ac.in,
girish.kulathumani2021@vitstudent.ac.in,
ganesh.narayanan@vit.ac.in

**Abstract.** Siren represents a pioneering research effort aimed at fortifying cybersecurity through strategic integration of deception, machine learning, and proactive threat analysis. Drawing inspiration from mythical sirens, this project employs sophisticated methods to lure potential threats into controlled environments. The system features a dynamic machine learning model for real-time analysis and classification, ensuring continuous adaptability to emerging cyber threats. The architectural framework includes a link monitoring proxy, a purpose-built machine learning model for dynamic link analysis, and a honeypot enriched with simulated user interactions to intensify threat engagement. Data protection within the honeypot is fortified with probabilistic encryption. Additionally, the incorporation of simulated user activity extends the system's capacity to capture and learn from potential attackers even after user disengagement. Overall, Siren introduces a paradigm shift in cybersecurity, transforming traditional defense mechanisms into proactive systems that actively engage and learn from potential adversaries. The research strives to enhance user protection while yielding valuable insights for ongoing refinement in response to the evolving landscape of cybersecurity threats

## 1 Introduction

The ongoing transformation of cyber threats underscores the critical need for adaptive and innovative defense strategies. Confronted with an expanding attack surface and increasingly sophisticated threat vectors, the traditional reactive measures often prove inadequate. Recognizing this imperative for state-of-the-art cybersecurity solutions, this research introduces the Siren system as an inventive and proactive defense mechanism. By integrating deception, machine learning, and real-time threat analysis, Siren seeks to contribute to the ongoing discourse on cybersecurity, offering insights and solutions tailored to the dynamic challenges posed by the ever-changing threat landscape.

Reactive strategies, while foundational, inherently operate in response to known threats, leaving organizations vulnerable to novel and evolving attack vectors. The escalating sophistication of cyber threats necessitates a paradigm shift toward proactive defense mechanisms. Proactive approaches enable anticipation and mitigation of potential threats before they materialize, offering a more robust and adaptive defense posture. This shift is imperative to stay ahead of adversaries, addressing vulnerabilities in advance and minimizing the impact of unforeseen and rapidly evolving cyber threats. The research presented here underscores the critical need for proactive cybersecurity measures, introducing the Siren system as a pioneering solution designed to navigate the complexities of the contemporary threat landscape.

Existing defensive paradigms predominantly hinge on reactive methodologies, responsive to known threats but lacking preemptive capabilities against unforeseen risks. The motivation to bridge this gap stems from the imperative to cultivate a cybersecurity approach that is not merely responsive but anticipatory, affording organizations the ability to proactively thwart potential threats. By introducing a proactive system, the Siren project aspires to revolutionize cybersecurity protocols, minimizing the lag time between threat emergence and defensive response. This research endeavors to address the critical need for anticipatory cybersecurity measures, aiming to contribute to the ongoing discourse in the field and fortify organizations against the ever-evolving landscape of cyber threats.

The aim is to create an environment where attackers are lured into controlled spaces by simulating vulnerabilities, facilitating an analysis of their behaviors. This unique approach seeks to leverage the power of deception to proactively engage with and understand potential threats, marking a departure from conventional cybersecurity methodologies. This involves the seamless integration of deception, adaptive machine learning, and real-time threat analysis within the Siren system. The research seeks to contribute empirically validated insights into the efficacy of proactive cybersecurity, examining its potential to preemptively identify and neutralize emerging threats.

Furthermore, the project aims to enhance the current understanding of adversary behaviors by meticulously analyzing their responses within controlled environments. By achieving these objectives, the Siren project aspires to establish a novel benchmark in cybersecurity practices, redefining the industry's approach to threat mitigation and contributing substantively to the evolution of proactive defense strategies. The foundational strategy of the Siren system involves a cohesive integration of deception, adaptive machine learning, and real-time threat analysis. This approach leverages the dynamic interplay between these core components to create a proactive defense mechanism. The Siren system is meticulously designed to anticipate and respond to emerging cyber threats, offering a comprehensive solution that transcends the limitations of reactive cybersecurity approaches.

Distinctive attributes define the Siren system, setting it apart in the realm of cybersecurity. An adaptive machine learning model forms the cognitive backbone, ensuring continuous evolution in response to emerging threats. The inclusion of a link monitoring proxy adds an additional layer of scrutiny, while the honeypot, fortified with probabilistic encryption, creates a secure and controlled environment. Furthermore, the incorporation of simulated user activity enhances the system's capacity to engage and learn from potential threats, marking a paradigm shift in cybersecurity strategies.

The envisioned outcomes of the Siren research project are twofold. Firstly, the system aims to elevate user protection by proactively identifying and mitigating potential threats. Through empirical validation and real-world application, Siren endeavors to demonstrate the efficacy of its approach in enhancing overall cybersecurity resilience. Secondly, the research anticipates contributing valuable insights to the broader cybersecurity domain. By analyzing adversary behaviors within controlled environments, Siren aims to provide nuanced perspectives that contribute substantively to the ongoing discourse on proactive defense strategies.

This research paper unfolds systematically, beginning with an "**Introduction**" that establishes the context of the current cybersecurity landscape and underscores the importance of proactive defense mechanisms. The subsequent section "**Related Works**" provides a detailed overview of the Siren system's approach, elucidating its core components and their integration. A dedicated segment ("**Proposed Methodology**") highlights the unique features of Siren, emphasizing the adaptive machine learning model, link monitoring proxy, honeypot with probabilistic encryption, and simulated user activity. The "**Experimentation and Results**" section provides a glimpse into the expected impact of the research on user protection and its contribution to cybersecurity insights. Finally, the paper concludes with a summary and recommendations for future research in the sections "**Limitations and Future work**" and **"Conclusion"**, providing a comprehensive roadmap for readers to grasp the significance and implications of the Siren project.

## 2    Related Works

In this paper, the possibilities of imbuing machine learning and probabilistic encryption with one another to strengthen security are explored. The current state of honeypots is that they are currently used as a secondary method to firewalls as a largely reactionary method according to Nawrocki et al. [1].

Honeypots, even if interacted with, generally just serve as a deterrent for attackers to waste time but can become a vulnerability if exposed for too long or largely interacted with. A honeypot design can be loosely based off a scanning tool, a fingerprinting device, and a few other tools from Kuwatly et al. [2]. The dynamic honeypot design offers several key advantages that make it a robust solution for network security. Its flexible data collection methods, combining both active (Nmap) and passive (P0f, Snort) fingerprinting, allow administrators to adapt to different network architectures. However, the design suffers from notable disadvantages that could impact its effectiveness. Performance issues include lengthy scanning times (up to 1300s for Linux systems) and high resource consumption during active scanning, which could strain network resources. The design's high resource requirements and need for frequent database maintenance add to its operational overhead.

Far more novel methods such adding an additional layer with user reconfirmation from Li and Schmitz [3] provide more security even with relatively prolonged interactions as well with very minimal if any leakage at all. This not only encourages the attacker to interact with the honeypot entity as a whole but can in fact keep the attacker from exploiting the vulnerability detected altogether. But, the framework faces notable challenges that could limit its widespread adoption and effectiveness. The complexity of implementation requires substantial server-side modifications and incurs additional costs for financial institutions. Therefore, the framework's effectiveness against increasingly sophisticated phishing techniques remains an open research question that requires extensive testing and refinement.

The main feature that makes honeypots a weak point is their easy access and vulnerability due to features such as cryptanalysis and other attacks performed on encryptions. However, the encryptions themselves can be strengthened if made probabilistic as put forward by Benaloh [7]. The probabilistic encryption scheme can be made using the standard encryption function of input and cipher but with a third variable that is random in nature.

Harn and Kiesler [6] presented a fairly efficient probabilistic scheme based on the concept of using a randomized variable to simplify functions such as Rabin's 4-1 scheme with randomized variables to ensure that the complexity of the encryption is retained but there is gain in speed. A probabilistic scheme does however pose the challenge of choosing the random variable as well and that in itself would need to be addressed in its entirety.

Lastly, deep learning frameworks to further protect the honeypot without making changes to the vulnerability itself can also be used. The model, according to Tang et al., can extract most of the detail using RNN-GNU models for decently engineered precision and some speed as well [4]. But the proposed model considers only the first 200 characters of each URL for phishing indicators, leading to potential loss of critical information for URLs exceeding this length, which may affect detection accuracy. The current review mechanism relies on predefined rules such as remote IP addresses, client information, and submission frequency which can be exploited by attackers to bypass detection.

Xuan, Nguyen and Tisenko [5] also further present that the URL can be sent to the machine learning model from the website itself to give near instant recognition to the user and administrator as part of the whole security of the organization. However the use of machine learning techniques such as SVM and Random Forest result in lower generalization possibilities than DL models can provide. Matin further enhances this using the honeypot itself to lure a phisher to come to the vulnerability it creates [8]. However, the problems of generalization are somewhat mitigated with the use of k-fold cross validation.

Gajek et al. [9] proposed a methodology involving the Phoneypot using falsified tokens to send back to the collection server in order to fool an attacker into thinking that they have in fact reached the honeypot. The system however has a weakness for transactions which could result in the phisher knowing that they are in the honeypot itself. Furthermore, content aware attacks leave the phoneypot extremely vulnerable to leakage of sensitive user data.

Osho et al. [11] conducted a comparative evaluation of phishing URL detection techniques, analyzing 35 classification algorithms, including Random Forest, decision trees, neural networks, and Bayesian models, on multiple publicly available datasets. Their study reaffirmed the effectiveness of machine learning-based approaches in identifying phishing URLs based on inherent URL characteristics. Random Forest consistently outperformed other classifiers in terms of accuracy, precision, recall, and

F-measure, demonstrating its robustness across different datasets. The findings align with prior research, which underscores the increasing sophistication of phishing attacks and the ongoing need for improved detection methodologies to enhance cybersecurity and user protection.

Yet none of the papers have managed to address the consistent vulnerability that the honeypot creates with the prolonged interaction of the phisher. Detection alone may not be entirely feasible on its own.

## 3 Proposed Methodology

### 3.1 Dataset and Feature Extraction

The dataset used for the project was a malicious links dataset sourced from Kaggle containing four different classes of URLs with 640,000 unique data points.

The dataset consists of two main columns, one for the specific URLs and the type of URL that they are. There are four main types of URLs in the dataset – benign, defacement, phishing and malware. The objective here is to simply classify the data based on the given link after extracting all possible features from it.

Features are extracted based on a few different categories from the URL itself. Firstly, the **protocol**, **host**, **path**, **parameters**, **query** and **fragment** are all separately extracted from the URL itself and placed into different columns of the dataset. This allows us to parse each component down to its most basic units and decide which features can be retained and which do not need retention later on. Any URL that does not have such features like query or fragment will be stored as null values.

There is a clear imbalance in features with a majority of the URLs provided belonging to the benign category. Even within the malevolent links, the majority of links belong to the category of defacement with phishing having a few lesser and barely any belonging to malware. The distribution for the data is visualized in Fig. 1.

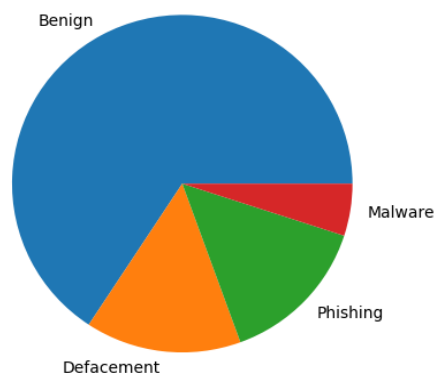

**Fig. 1.** A pie chart showing the data distribution of the links in the given dataset

## 3.2 URL Features

Columns with predominantly null values are often redundant and do not contribute significantly to the predictive or analytical power of a model. Therefore, such columns are systematically removed during the preprocessing stage to ensure the model remains efficient and accurate. In this particular case, the columns identified and removed due to their high percentage of null values were parameters, query, and fragment. These columns lacked meaningful data and thus were deemed unnecessary for further analysis or model development.

Once these columns were eliminated, attention shifted to other columns containing null values. These columns were carefully extracted for closer inspection, and appropriate measures were taken to address the missing values. Typically, this involves strategies such as removing rows with null values, imputing them with statistical measures like the mean, median, or mode, or applying domain-specific knowledge to handle the gaps effectively. In this context, null values were removed outright to maintain data integrity.

Following this, regular expressions were employed to extract valid IP addresses from the dataset. This process involved identifying and isolating various formats of IP addresses, including IPv4, IPv4 with port numbers, IPv4 in hexadecimal representation, and IPv6 addresses. Regular expressions, being a powerful tool for pattern matching, ensured that only valid IP addresses were captured. This rigorous cleaning and transformation step was critical in preparing the dataset for subsequent modeling and analysis, ensuring both accuracy and consistency.

To enhance the analysis of URLs, those containing special symbols were extracted and stored in designated columns. This allowed for the identification of any predominant symbols that might have been used in the construction of the URLs, which could serve as indicators of specific patterns or behaviors. Additionally, the length of each URL was calculated and retained as a feature to analyze its correlation with potential add-on domains often associated with malicious websites. Shortened URLs generated through popular services such as Bit.ly, TinyURL, and others were also carefully considered. These links were not excluded from the dataset but were categorized separately to ensure they could be appropriately classified.

Several URL characteristics were further analyzed to provide additional insights. Features such as digit counts, email addresses, letter counts, vowel counts, and top-level domain (TLD) counts were examined in detail. After careful evaluation, TLD counts were retained as a key feature, given their relevance in distinguishing between legitimate and malicious URLs. This decision ensured that unnecessary noise in the dataset was minimized while preserving critical information.

Furthermore, the possibility of server-client information being embedded within URLs was also analyzed as a classification feature. This step was particularly significant for identifying fake, malicious replicas of legitimate websites. By scrutinizing server-client details, the approach aimed to detect subtle indicators that might reveal fraudulent activity. Together, these features and analytical steps formed a robust framework for improving classification accuracy and identifying patterns associated with potentially harmful URLs.

### 3.3 Path Features

The first feature extracted from the path was the presence of special symbols, following the same approach used for analyzing URLs. These symbols were identified and analyzed to determine their frequency and potential patterns, as they might indicate specific behaviors or purposes related to the path structure. Additionally, other key characteristics were extracted from the path to provide a more comprehensive analysis.

Features such as the length of the path, digit counts, letter counts, and vowel counts were calculated and recorded. The length of the path served as an essential feature to identify patterns associated with longer or shorter paths, which could correlate with specific functionalities or malicious intentions. Digit counts provided insight into numeric patterns, while letter and vowel counts contributed to understanding the overall composition of the path. Together, these features helped uncover critical patterns and enhance the ability to classify and analyze path data effectively.

### 3.4 Head Features

In this case as well, special symbols were extracted as a key feature to analyze their occurrence and potential significance. Alongside special symbols, other important characteristics such as the length of the header, digit counts, letter counts, and vowel counts were also extracted. The length provided insight into the complexity or simplicity of the path structure, which could correlate with specific behaviors or functionalities. Digit counts helped identify numeric patterns, while letter and vowel counts offered a better understanding of the overall textual composition. These features collectively contributed to a more detailed analysis and improved the classification and detection process.

### 3.5 Model

The model that is used is a basic perceptron for classification using a 68 dimensional input to make sure all the features extracted thus far can be input into the neural network. The hidden layers have a dimension of 300, 200 and 100 respectively.

To ensure the highest possible chance of generalization, two dropout layers to the model with a probability of 0.2 and a batch normalization layer have been added before it is fed to the output, assuming a softmax activation. The hidden layers are all activated using the Scaled Exponential Linear Unit or SELU [12] activation illustrated in Eq. (1).

$$SELU(x) = \alpha x \text{ if } x \geq 0$$
$$SELU(x) = \alpha \gamma (\exp(x) - 1) \text{ if } x < 0 \quad (1)$$

The output is passed through a softmax layer in order to classify it into one of the four classes. The architecture of the model is illustrated in Fig. 2.

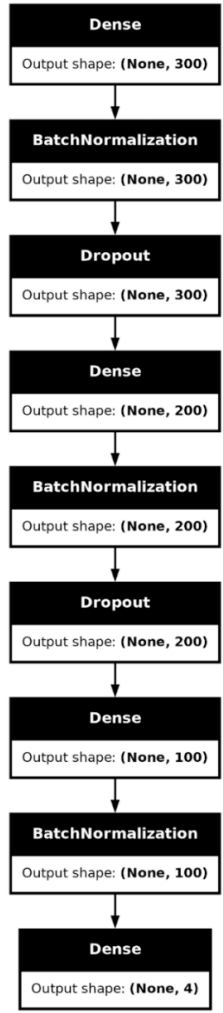

**Fig. 2.** The model architecture in sequential format with each layer described

### 3.6 Honeypot

The honeypot is made using a standard architecture with the intention of giving the attacker a method of trying to access the system under the guise of being a dummy that simply serves no purpose in and of itself other than bait. The honeypot needs to be able to sustain itself to make sure that the attacker remains within the honeypot itself for as long as the attacker's belief on the fact that the system is a real vulnerable system that can be exploited can be maintained [10].

The honeypot itself also needs to ensure a sort of detachment from the main systems to make sure the attacker cannot use the honeypot as a breaching point very easily. The main methodology suggested here is the use of a one-way connection to connect the mainframe to the honeypot so that some communication is possible.

The one-way further allows encryption commands to be sent to the honeypot so that the files can be encrypted before access just to make sure the attacker cannot gain access to the files very easily, even if they are dummy files so as to waste the attacker's time and effort.

This further enables the study of the attacker's patterns if monitoring the honeypot is added so that the attack patterns can be learnt with time as the attacker is executing them.

The main stalling feature that will be used is a probabilistic key generation algorithm on the RSA encryption to prevent cryptanalysis of any sorts between texts. The major architecture of the honeypot is shown in Fig. 3.

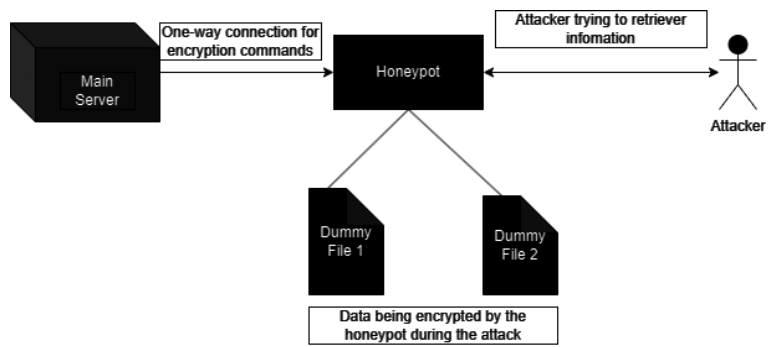

**Fig. 3.** Overall honeypot architecture

### 3.7 Probabilistic RSA Encryption

Probabilistic RSA encryption is an enhancement of the traditional RSA encryption algorithm that introduces randomness into the encryption process to bolster security against certain types of attacks. RSA encryption, named after its inventors Rivest, Shamir, and Adleman, is an asymmetric encryption scheme widely used for securing communications and data transmission [6].

In traditional RSA encryption, plaintext is encrypted using the recipient's public key, and the resulting ciphertext can only be decrypted with the corresponding private key. However, this basic scheme is susceptible to certain vulnerabilities, particularly when encrypting small or predictable plaintexts. Probabilistic RSA encryption addresses these vulnerabilities by incorporating randomness during encryption.

The fundamental principle behind probabilistic RSA encryption is to introduce randomness into both the plaintext and the encryption process itself. This randomization helps thwart attacks that exploit regularities or patterns in the plaintext data. Unlike deterministic encryption schemes, where the same plaintext always encrypts to the same ciphertext, probabilistic encryption ensures that each encryption operation produces a different ciphertext, even for the same plaintext.

The probabilistic RSA encryption uses the OpenWeather API to take the weather data of multiple cities and uses them to generate keys.

The process of probabilistic RSA encryption without padding can be summarized as follows:

**Key Generation.**
- Take 2 random numbers between 1 and the total number of cities being considered.
- Find the city associated with the first number.
- Take the maximum and minimum temperature for that day in Kelvin. Find the difference of the two temperatures.
- Multiply it by an arbitrary number x to get a value and find the nearest prime to that value.
- Repeat the above process to get the two primes needed for the encryption.
- Once the primes have been received, product, totient of the product, public and private key can be generated.
- The public key and product are sent to the sender of the ciphertext.

**Sender**
- The sender receives the public key and encrypts his plaintext using modular exponentiation by raising the text to the value of the public key modulo product.
- The ciphertext can be sent to the receiver.

**Receiver.**
- The receiver can decrypt the ciphertext by raising to the value of the private key modulo product.

Probabilistic RSA encryption provides several security advantages:
- *Randomization:* By introducing randomness into the encryption process, probabilistic RSA encryption ensures that the same plaintext encrypts to different ciphertexts each time, thwarting attacks based on patterns or repetitions in the data.
- *Resistance to chosen-ciphertext attacks:* Probabilistic encryption schemes make it difficult for attackers to exploit interactions with the encryption oracle to obtain information about the plaintext.
- *Enhanced security for small or predictable plaintexts:* Randomization helps mitigate vulnerabilities associated with encrypting plaintexts with low entropy or predictable structure.

Overall, probabilistic RSA encryption enhances the security of RSA by introducing randomness into the encryption process, making it more resistant to various cryptographic attacks. While the basic RSA encryption scheme provides robust security for many applications, probabilistic encryption techniques offer additional protection against specific threats, making them valuable tools in securing sensitive data and communications.

## 4  Experimentation and Results

The experimentation was done on Python 3.11 in a Jupyter Notebook with the tensorflow library and Cowrie to simulate the honeypot in order to test its working [10].

The dataset was split into a 75:15:10 split for training, testing and validation respectively. All data splits were done using the stratified k-fold algorithm with 5 folds used to make sure that the data splits had relatively equal occurrences of each class and no class was missing from any one data split.

All weight and kernel initializations are done using Lecun to make sure that samples are drawn from a standard Gaussian curve of mean 0 and standard deviation 1. The loss function used for this is a Categorical Crossentropy loss function with a Nesterov-accelarated Adaptive Moment Estimation optimizer (NAdam) [13] to ensure adequate change in learning rate depending on the time. The metrics used for the model are Recall and Loss.

The model was also early stopped at the best epoch and saved on callback to make sure that the best point of training and validation can be taken for evaluating links. Here, the model was taken from epoch number 28 due to its similarity in training and validation recall. The graphs for the model are shown in Fig. 4.

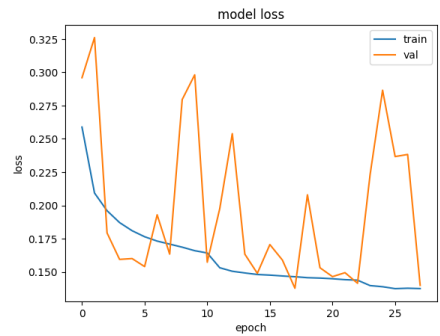

(a)

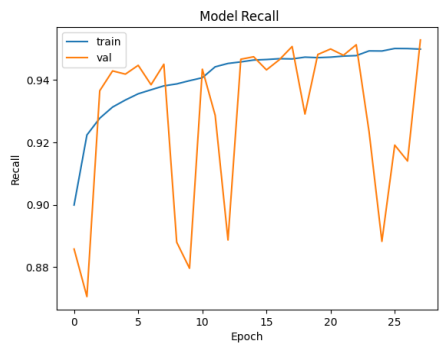

(b)

**Fig. 4.** Model results on training and validation data for (a) Loss and (b) Recall

Overall results for the model show similar results on testing data as well with the metrics illustrated in Fig. 5.

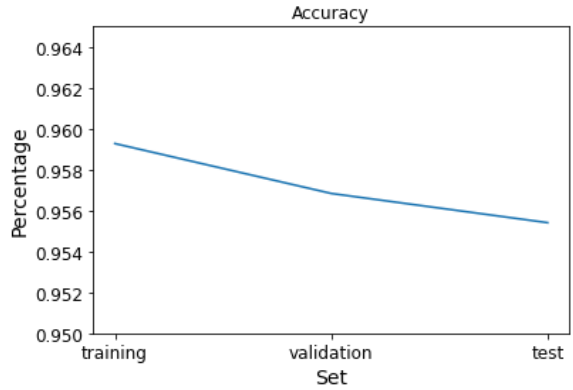

(a)

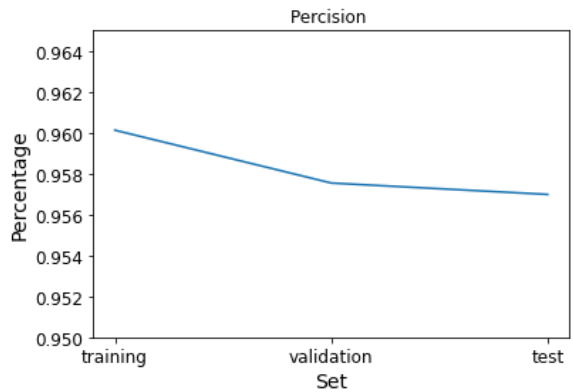

(b)

**Fig. 5.** Model metrics of loss and accuracy across the data split for (a) Accuracy and (b) Precision

The overall metrics have been tabulated in Table 1.

**Table 1.** Overall metrics of the model across the data splits

| Metric | Training | Validation | Testing |
| --- | --- | --- | --- |
| Loss | 0.076 | 0.079 | 0.077 |
| Recall (%) | 96.2 | 95.8 | 96.9 |
| Precision (%) | 96.1 | 95.8 | 95.8 |
| Accuracy (%) | 95.9 | 95.8 | 95.8 |

The model has been ensured to the extent that no overfitting has taken place in any of the datasets with recall differences being limited to around 0.0028 for training and validation recall and 0.0032 for training and validation loss. The model can now be used for detection of fraudulent or malevolent links being clicked within the framework so as to activate the honeypot and check if any intruder has entered within the given time frame.

## 5 Limitations and Future Work

Major improvements to this field can be made to the siren framework by adding more improvements to the honeypot to make studying attacker patterns and signatures far easier and with lower risk. Furthermore, the scalability of such a system is something to be addressed due to the inference time of the machine learning model along with more robust key generation schemes or encryptions that could hold off any attacker for far longer while still being fairly light and easy to deploy. The model also constantly needs to be retrained and updated on newer attack frameworks and links which leads to downtime. Lastly, a backup or counter-deception method added as a failsafe could make Siren a far more complete safety method.

There's a good chance that this research will significantly advance cybersecurity. User protection will be strengthened by the empirical evaluation of Siren's effectiveness in proactively recognizing and reducing potential threats. Moreover, studying how adversaries behave in controlled settings will provide insightful information that will significantly influence the creation of proactive defensive plans in the future.

## 6 Conclusion

This paper presents research on Siren, a proactive cybersecurity protection mechanism that combines machine learning, deception, and real-time threat analysis. Siren makes it easier to analyze prospective attacker activity and gives the system the ability to proactively stop threats by deliberately luring them into controlled surroundings. The fundamental principle of Siren is its capacity to detect and react to new threats, presenting a major departure from reactive cybersecurity techniques.

The results of this study highlight how urgent it is to advance beyond reactive cybersecurity strategies. Organizations may fortify their defenses and remain ahead of the constantly changing threat landscape by implementing proactive solutions like Siren. Subsequent research paths involve the extensive implementation and assessment of Siren in real-world settings, in addition to constant improvement of the system's elements to guarantee its effectiveness against new and complex dangers from cyberspace. Siren clears the path for a day where cybersecurity is strengthened and proactive rather than reactive, protecting digital assets and reducing the dangers associated with persistent cyberattacks.